\documentclass[11pt, twoside]{article}
\usepackage{a4wide}
\usepackage{graphicx}

\begin{document}
\pagestyle{myheadings}
\markboth{WLPE'01}{On the Design of a Tool for Supporting the Construction of Logic Programs}

\title{On the Design of a Tool for Supporting the Construction of Logic Programs\footnote{In A. Kusalik (ed), Proceedings of the Eleventh Workshop on Logic Programming Environments (WLPE'01),
December 1, 2001, Paphos, Cyprus. COmputer Research Repository (http://www.acm.org/corr/), cs.PL/0111041;
whole proceedings: cs.PL/0111042.}}

\author{Gustavo A. Ospina \and Baudouin Le Charlier\\
D\'epartement d'Ing\'enierie Informatique, Universit\'e Catholique de Louvain\\
Place Sainte Barbe 2, B-1348 Louvain-la-Neuve, Belgium\\
{\tt \{gos,blc\}@info.ucl.ac.be}}
\date{}
\maketitle

\begin{abstract}
Environments for systematic construction of logic programs are needed in the academy as well as in the industry. Such environments should support well defined construction methods and should be able to be extended and interact with other programming tools like debuggers and compilers. We present a variant of the Deville methodology for logic program development, and the design of a tool for supporting the methodology. Our aim is to facilitate the learning of logic programming and to set the basis of more sophisticated tools for program development.
\end{abstract}

\section*{Introduction}
Logic Programming is a suitable paradigm for supporting methods and techniques of software engineering, as shown in \cite{ciancarini}. However, there are relatively few attempts to bring software support to logic program construction.

The interest in having computer tools to help logic program development is twofold: tools can help novice programmers and students to get skills in the programming paradigm; on the other hand, in an industrial context, tools can make logic programming more widely used in the development of large software systems.

Tools are usually designed to give (semi)automatic support to methodologies of program development. Well defined methodologies help the programmer in the construction process, starting from a specification, to get a program which is correct with respect to that specification. There are several methods to support the construction of logic programs. We can mention Deville's Methodology \cite{deville}, Stepwise Enhancement \cite{jain} with its higher order extensions \cite{naish} and Synthesis from Schemas \cite{flener}.

In this paper we present some notes on the design of a tool to support Prolog program develoment based on a variant of Deville's methodology \cite{deville}. This variant adds a type system and the notion of typed logic description, which allows a possible adaptation of the methodology for program construction in typed logic languages like Mercury \cite{mercury}. Our tool is designed to help students in their learning of logic programming, and to bring another chance in the development of CASE tools for this paradigm.

The paper is structured as follows. The next section describes the variant to Deville's Methodology for logic program development. Section 2 gives details on the design of a tool to support the methodology. Section 3 will conclude and give perspectives for further work.

\section{Methodology for Logic Program Construction}

The original methodology proposed by Deville in \cite{deville} consists of three steps:

\begin{enumerate}
\item Elaboration of a specification
\item Construction of a logic description
\item Derivation of a logic procedure
\end{enumerate}

The result of the application of the methodology is the set of clauses for a logic procedure {\em p/n}, this is, clauses with the same head {\tt p(X1,...,Xn)}. When the methodology is correctly applied, the logic procedure is correct with respect to its specification.

Our new contribution to the original methodology is the division in the second step:

\begin{enumerate}
\item Elaboration of a specification
\item
  \begin{enumerate}
  \item {\em Construction of a typed logic description}
  \item {\em Conversion of the typed logic description into a non-typed logic description}
  \end{enumerate}
\item Derivation of a logic procedure
\end{enumerate}

We add a simple type system to support the construction of typed logic descriptions. Such typed logic descriptions are more suitable to be translated to typed logic languages.

\subsection{Elaboration of a specification}
Specifications follows the template proposed by Deville and revised in the FOLON project \cite{henrard}, with slight modifications. The specification template is shown in Fig. 1.

\begin{figure}
\begin{tabular}{p{12cm}}
\hline\\
{\bf procedure p(X$_1$,...,X$_n$)}\\
\\
{\em Types: }\\
X$_1$ : type$_1$\\
.\\
.\\
.\\
X$_n$ : type$_n$\\
\\
{\em Relation:}\\
Description of a relation between the parameters X$_1$,...,X$_n$\\
\\
{\em Directionalities:}\\
Description of directionalities, with mode, determinism and no sharing information\\
\\
{\em External conditions:}\\
Description of the external environment preconditions and side effects\\
\hline
\end{tabular}
\caption{General specification template}
\end{figure}

We use the same approach of the methodology of having types as sets of ground terms. However, we also define a type system to restrict the sets that can be considered as a type. The type system is described in the next section.

A directionality for a procedure {\em p/n} has the form:

\begin{center}
($M_1,...,M_n$): $Mult$: $No\_sh$
\end{center}

Where:

\begin{itemize}
\item $M_i = In_i \mapsto Out_i$,  $In_i, Out_i \in$ \{{\tt ground, ngv, var, novar, gv, noground, any}\}. A singleton $Mode$ abbreviates $Mode \mapsto Mode$.
\item $Mult =~<Min-Max>$ where Min, Max $\in N \cup \{*, \infty\}$
\item $No\_sh$ is a set of pairs $(i,j), i \neq j$ and $i,j \in 1..n$
\end{itemize}

The $M_i$ bindings are usually called the {\em modes} of the procedure. We decided to represent them in the form $In_i \mapsto Out_i$ to make them more intuitive and closer to mode declarations in Mercury.

$Mult$ is the {\em determinism} associated with modes $M_i$. It estimates the lower and upper bounds\footnote{A bound '*' indicates a fixed but unknown number of answer substitutions, whereas '$\infty$' indicates infinite answer substitutions} in the number of answer substitutions of the procedure when it is invoked with a entry substitution respecting to the $In_i$ in the mode declaration. Fig. 2 shows a correspondance between mode and determinism declaration in our specification template and the ones used by Mercury.

\begin{figure}
\begin{center}
\begin{tabular}{*{2}{p{6.5cm}}}
\begin{tabular}{|p{2.2cm}|p{3.8cm}|}
\hline
Mercury & $<Min-Max>$\\
determinism & determinism\\
\hline
{\tt det} & {\scriptsize $<1-1>$}\\
{\tt semidet} & {\scriptsize $<0-1>$}\\
{\tt nondet} & {\scriptsize $<0-*> <0-\infty>$}\\
{\tt multi} & {\scriptsize $<1-*> <1-\infty> <*-*>$}\\
{\tt failure} & {\scriptsize $<0-0>$}\\
{\tt erroneous} & {\scriptsize $<1-0>$}\\
\hline
\end{tabular}
&
\begin{tabular}{|p{1.5cm}|p{3.5cm}|}
\hline
Mercury & $In \mapsto Out$\\
mode & directionality\\
\hline
{\tt in} & $ground \mapsto ground$\\
{\tt out} & $var \mapsto ground$\\
{\tt di} & $ground \mapsto ground$\\
{\tt uo} & $var \mapsto ground$\\
{\em User} & $any \mapsto any$\\
{\em defined} & \\
\hline
\end{tabular}
\end{tabular}
\end{center}
\caption{Comparison with Mercury modes and determinism}
\end{figure}

$No\_sh$ is a set of pairs $(i, j)$, each pair means that parameters $T_i$ and $T_j$ do not share variables. It can be seen as a partial description of a reflexive and symmetric relation on the set of parameters.

\subsection{Construction of a Typed Logic Description}

We recall the definition of a (non typed) Logic Description according to Deville \cite{deville}:

\paragraph{Definition 1.}
A {\em (non typed) Logic Description} is a closed first order logic formula of the form $$\forall X_1 ... \forall X_n \bullet p(X_1,...,X_n) \Leftrightarrow Def$$ Where {\em Def}, the {\em definition part} of the logic description, is a closed first order logic formula. (Non typed) logic descriptions are denoted as {\em LD(p/n)}.

For the sake of simplicity, we omit the universal quantifiers in $p(X_1,...,X_n)$ and assume the free variables in {\em Def} as existentially quantified.
\\
\\
In \cite{deville} the notion of correction of a logic description with respect to its specification is defined, and some construction techniques based on {\em structural induction} are proposed, in order to build the definition part of the logic description. The resulting logic description is ``correct by construction''.

The structural induction technique to build the definition part of {\em LD(p/n)} produces a logic description of the form
$$
p(X_1,...,X_n) \Leftrightarrow (C_1 \wedge F_1)\\
\vee \\
.\\
.\\
.\\
\vee (C_m \wedge F_m)
$$

In the following, we call $Term$ to be the set of all the ground terms.

The definition part of the logic description is obtained after following these steps:

\begin{itemize}
\item
Choose {\em induction parameters} on parameters X$_1$,...,X$_n$, denoted I$_1$,...,I$_p$ ($p \leq n$)
\item
Find a {\em well-founded relation} over the cartesian product $Term^p$. Such relations are found with the help of type information.
\item
Define the {\em structural forms} $C_1,...,C_m$ of induction parameters I$_1$,...,I$_p$, according to the well-founded relation.
\item
Build the {\em structural cases}. For each $C_i$, build the corresponding $F_i$, which is a logic formula that makes true the predicate {\em p} whenever $C_i$ is true. $F_i$ can have occurrences of {\em p}.
\item
Add type-checking literals when needed.
\end{itemize}

Variants of this technique include tupling generalization and introduction of accumulators \cite{deville}.

We will see that these techniques are still applicable for typed logic description, with the advantage of a better handling of type-checking literals. This requires the definition of a type system.

\subsubsection{Type system}
According to the original methodology, a type is simply an arbitrary set of ground terms. We propose a type system which describes such sets in a more structured way. Our type system is enoughly expressive to describe several useful types, altough it lacks some desirable features of modern type systems like parametric polymorphism.

The set $Term$ is the only type which contains all the other types.

A type is described with a non typed logic description over an unary predicate, which has this form: $$type(X) \Leftrightarrow TypeDef$$

$TypeDef$ cannot be any logic formula. It must have one of these forms:

\begin{itemize}
\item
$C_1 \vee ... \vee C_m$ where each $C_i$ is a conjunction $X = type\_cons(X_1,...,X_l) \wedge type_1(X_1) \wedge ... \wedge type_l(X_l)$. To abbreviate we can omit the occurrences of $term(X)$ for any variable $X$, since it is equivalent to $true$. Note that the name of the type to be defined can occur in the $C_i$. A type defined in that way is called {\em recursively defined}.
\item
$another\_type(X)$. In this way one describes a {\em type equivalence}. It means that the definition of $type$ is equal to the definition of $another\_type$. Types cannot be mutually recursive. For instance it is not allowed to have two types $t_1$ and $t_2$ defined as:
\begin{center}
\begin{tabular}{c}
$t_1(X) \Leftrightarrow t_2(X)$\\
$t_2(X) \Leftrightarrow c(Y) \vee t_1(Y)$
\end{tabular}
\end{center}
\end{itemize}

Describe enumeration types is easy with this syntactic sugar: $$enum\_type(X) \Leftrightarrow X \in \{atom_1,...,atom_m\}$$ is equivalent to $$enum\_type(X) \Leftrightarrow X = atom_1 \vee ... \vee X = atom_m$$

These are some examples of types:

\begin{itemize}
\item $fruit(X) \Leftrightarrow X \in \{orange, apple, banana, pineapple, strawberry\}$
\item $nat(X) \Leftrightarrow X = zero \vee (X=s(N) \wedge nat(N))$
\item $list(X) \Leftrightarrow (X = empty\_list)~\vee~(X = cons\_list(H,T) \wedge list(T))$
\item $nat\_list(X) \Leftrightarrow (X = empty\_list)~\vee~(X = cons\_list(H, T)~\wedge~nat(H)~\wedge~nat\_list(T))$
\item $nat\_set(X) \Leftrightarrow nat\_list(X)$
\end{itemize}

In the following, we will call $Type$ the set of all the names of types.

\subsubsection{Typed Logic Descriptions}

Typed logic descriptions use a typed first order logic, where the only difference with classic first order logic is the addition of type information in quantifiers. For instance, instead of writing $\forall X \bullet p(X)$ we write $\forall X \in T \bullet p(X)$, where $T \in Type$

\paragraph{Definition 2.}
A {\em Typed Logic Description} is a closed typed first order logic formula of the form $$\forall X_1 \in type_1 ... \forall X_n \in type_n \bullet p(X_1,...,X_n) \Leftrightarrow TypedDef$$ Where $type_1,...,type_n$ and the type names in all the quantifiers found in $TypedDef$ belongs to $Type$. {\em TypedDef}, the {\em definition part} of the logic description, is a closed typed first order logic formula. Typed logic descriptions are denoted as {\em TLD(p/n)}.

Once again, we can make somme abbreviations. We omit the universal quantifiers to annotate the parameters with their types. Free variables in $TypedDef$ are assumed to be existentially quantified and belonging to the type $Term$. The general form of a typed logic description is then written as: $$p(X_1:type_1,...,X_n:type_n) \Leftrightarrow TypedDef$$

Typed logic descriptions are close to declarative descriptions without functions and higher order defined in \cite{baldan}, which are the base to adapt the Deville methodology to the development of programs in Mercury\footnote{Some of the main features of Mercury are strong types, facilities for higher order programming and functional notation \cite{mercury}.}.

In our variant of the methodology, firstly we construct a typed logic description. We can use the same technique of structural induction for non typed logic descriptions. The advantages of using typed logic descriptions instead of non typed are these:

\begin{itemize}
\item Programmer must find a well founded relation over $type_1 \times ... \times type_p$, rather than finding it over $Term^p$.
\item Structural forms are suggested by type definitions.
\item Programmer does not have to add manually type-checking literals.
\end{itemize}

The typed logic description constructed with this technique can be directly translated to typed languages like Mercury and Godel. See the section 1.4.

Type definitions can be also seen as skeletons \cite{sterling} on which can be applied techniques in order to derive a logic program.

\subsection{Conversion from a Typed Logic Description into a Non Typed Logic Description}

We are interested in the development of Prolog programs, which are not typed. In order to obtain this, the typed logic description constructed needs to be expressed in classical first order logic. This is made by converting the typed logic description into a non typed logic description.

It is required that the transformation preserves the declarative semantics of first order logic. We must define a notion of {\em equivalence} between a typed logic formula and a non typed logic formula.

\paragraph{Definition 3.}
A typed logic formula $T_F$ with free variables $X_1,...,X_n$ belonging to types $T_1,...,T_n$ is {\em equivalent} to a non typed logic formula $F$ with the same free variables $X_1,...,X_n$ if and only if:

\begin{itemize}
\item
If there is an instance $x_1,...,x_n$ of variables $X_1,...,X_n$ such that for some $i, (1 \leq i \leq n)$, $x_i$ is a term not belonging to the corresponding type $T_i$, then $F$ is false.
\item
For all the instances $x_1,...,x_n$ of variables $X_1,...,X_n$ such that $x_i$ is a term belonging to the type $T_i$, $F \equiv T_F$.
\end{itemize}

\paragraph{Definition 4.}
{\em (Transformation of a typed logic formula into a non typed logic formula)}. Let $F$ be a typed logic formula, with free variables $X_1,...,X_n$ of types $T_1,...,T_n$. Let  $check_F$ be the non typed formula $T_1(X_1) \wedge ... \wedge T_n(X_n)$.

$F^{nt}$ is the non typed logic formula resulting of the transformation showed in Fig. 3.

\begin{figure}
\begin{center}
\begin{tabular}{||c|c||}
\hline
\hline
$F$ & $F^{nt}$\\
\hline
\hline
$X_{i_1} = X_{i_2}$ & $X_{i_1} = X_{i_2} \wedge T_{i_1}(X_{i_1}) \wedge T_{i_2}(X_{i_2})$\\
\hline
$X_{i_1} = f(X_{i_2},...,X_{i_n})$ & $X_{i_1} = f(X_{i_2},...,X_{i_n}) \wedge T_{i_1}(X_{i_1}) \wedge ... \wedge T_{i_n}(X_{i_n})$\\
\hline
$\exists X \in T \bullet G$ & $\exists X \bullet T(X) \wedge G^{nt}$\\
\hline
$\forall X \in T \bullet G$ & $\forall X \bullet T(X) \Rightarrow G^{nt}$\\
\hline
$G \wedge H$ & $G^{nt} \wedge H^{nt}$\\
\hline
$G \vee H$ & $(G^{nt} \wedge check_H) \vee (H^{nt} \wedge check_G)$\\
\hline
$\neg G$ & $\neg G^{nt} \wedge check_G$\\
\hline
$G \Rightarrow H$ & $(\neg G^{nt} \wedge check_G \wedge check_H) \vee (H^{nt} \wedge check_G)$\\
\hline
$G \Leftrightarrow H$ & $(G^{nt} \Leftrightarrow H^{nt}) \wedge check_G \wedge check_H$\\
\hline
\hline
\end{tabular}
\end{center}
\caption{Transformation of a typed formula $F$ into a non typed formula $F^{nt}$}
\end{figure}

\paragraph{Definition 5.}
{\em (Transformation of a typed logic description to a non typed logic description)} Let {\em TLD(p/n)} be the typed logic description $$p(X_1:T_1,...,X_n:T_n) \Leftrightarrow Def$$ The resulting non typed logic description {\em LD(p/n)} is

\begin{center}
\begin{tabular}{c c l}
$p(X_1,...,X_n)$ & $\Leftrightarrow$ & $T_1(X_1)$\\
 & & $\wedge~...$\\
 & & $\wedge~T_n(X_n)$\\
 & & $\wedge~Def^{nt}$
\end{tabular}
\end{center}

Where $Def^{nt}$ is obtained by transforming the typed logic formula $Def$.

Type-checking literals are introduced in the transformation. Much of these literals are redundant and can be eliminated. This is obtained by analysing types and modes in the generated logic program.

\subsection{Derivation of a logic program}

A first version of a procedure in the implementation logic language is derived from the non typed logic description\footnote{when the target logic language is strongly typed, this derivation can be made directly from the typed logic description.}, following the syntactical transformation rules described in \cite{deville}. These rules are based in Clark completion.

The final version of the procedure is obtained after these steps:

\begin{itemize}
\item
Eliminating redundant type-checking literals
\item
Finding of a permutation of literals in the clauses that satisfy the specified directionalities
\item
Making language-dependent optimizations
\end{itemize}

The first two steps are carried out by a combined static analysis of types and directionalities in the program. It is possible that no permutation satisfying the directionalities can be found. In such case, there are two alternatives: change the specification of directionalities in a way that a permutation can be found, or ``split'' the procedure to handle each directionality separately.

Language-dependent optimizations for Prolog include cut introduction and forward and backward variable substitutions.

\subsection{Example}

We specify the procedure {\em max\_prefix/2}, which find the maximum of the sums of all the prefixes of an integer list.

\begin{center}
\begin{tabular}{p{12cm}}
\hline\\
{\bf procedure max\_prefix(L, M)}\\
\\
{\em Types: }\\
L : integer\_list\\
M : integer\\
\\
{\em Relation:}\\
M is the maximum of the sums of all the prefixes of L\\
\\
{\em Directionalities:}\\
(ground, var $\mapsto$ ground) : $<1-1>$\\
(ground, ground) : $<0-1>$\\
\hline
\end{tabular}
\end{center}

We will solve this problem using structural induction with introduction of accumulators, which is a particular case of computational generalization. A new generalized procedure {\em max\_prefix\_gen/3} is specified.

\begin{center}
\begin{tabular}{p{12cm}}
\hline\\
{\bf procedure max\_prefix\_gen(L, M, A)}\\
\\
{\em Types: }\\
L : integer\_list\\
M : integer\\
A : integer
\\
{\em Relation:}\\
M = M'+A, where M' is the maximum of the sums of all the prefixes of L\\
\\
{\em Directionalities:}\\
(ground, var $\mapsto$ ground, ground) : $<1-1>$\\
(ground, ground, ground) : $<0-1>$\\
\hline
\end{tabular}
\end{center}

We will use the structural induction technique with the generalized procedure.

\begin{itemize}
\item {\em Induction parameter:} L
\item {\em Well founded relation:} proper list suffix
\item {\em Structural forms:} $L=[~]$ and $\exists H \in Integer, T \in List \bullet L=[H|T]$
\item {\em Structural cases:}
\begin{itemize}
\item If $L=[~]$, trivially $M=A$.
\item If $L=[H|T]$, let $M'$ the maximum of the sums of prefixes of $T$, having accumulated $H+A$. M is the maximum between $H+A$ and $M'$.
\end{itemize}
\end{itemize}

The resulting typed logic description is

$max\_prefix\_gen(L:Integer\_list, M:Integer, A:Integer) \Leftrightarrow$
\begin{center}
\begin{tabular}{c l}
 & $L=[~] \wedge M=A$\\
$\vee$ & $\exists M' \in Integer~\bullet$\\
 & $L=[H|T]$\\
 & $\wedge~max\_prefix\_gen(T, M', H+A)$\\
 & $\wedge~max(H+A, M', M)$
\end{tabular}
\end{center}

After eliminating redundant type-checking literals we obtained the pure Prolog procedures presented in Fig. 4. We also present a possible implementation in Mercury in Fig. 5, using the correspondances in Fig. 2 to get the declarations of the procedure.

\begin{figure}
\begin{verbatim}
max_prefix_gen(L, M, A) :-
    L = [],
    M = A,
    integer(M).

max_prefix_gen(L, M, A) :-
    L = [H | T],
    plus(H, A, A1),
    max_prefix_gen(T, M1, A1),
    max(A1, M1, M).

% max_prefix as a particular case of max_prefix_gen
%
max_prefix(L, M) :-
    max_prefix_gen(L, M, -infinite).
\end{verbatim}
\caption{Prolog code for {\em max\_prefix\_gen/3} and {\em max\_prefix/2}}
\end{figure}

\begin{figure}
\begin{verbatim}
:- pred max_prefix_gen(integer_list, integer, integer).
:- mode max_prefix_gen(in, out, in) is det.
:- mode max_prefix_gen(in, in, in) is semidet.

max_prefix_gen(L, M, A) :-
(   L = [],
    M = A
;
    L = [H | T],
    max_prefix_gen(T, M1, A + H),
    max(A + H, M1, M)
).

% max_prefix is a particular case of max_prefix_gen
%
:- pred max_prefix(integer_list, integer).
:- mode max_prefix(in, out) is det.
:- mode max_prefix(in, in) is semidet.

max_prefix(L, M) :-
    min_int(X),
    max_prefix_gen(L, M, X).
\end{verbatim}
\caption{Mercury code for {\em max\_prefix\_gen/3} and {\em max\_prefix/2}}
\end{figure}

\section{Elements for a computational tool}

Our design of a computational tool for supporting the methodology is inspired by the FOLON environment \cite{henrard} and the prototype SPG \cite{ospina}.

One of our goals is the availability of the tool to interact with existing programming environments and to facilitate its extension with other methods and techniques of logic program development. In order to obtain this, each step of methodology is supported by one or many software components. Fig. 6 shows the basic component architecture of the tool.

\begin{figure}
\begin{center}
\includegraphics[width=8cm]{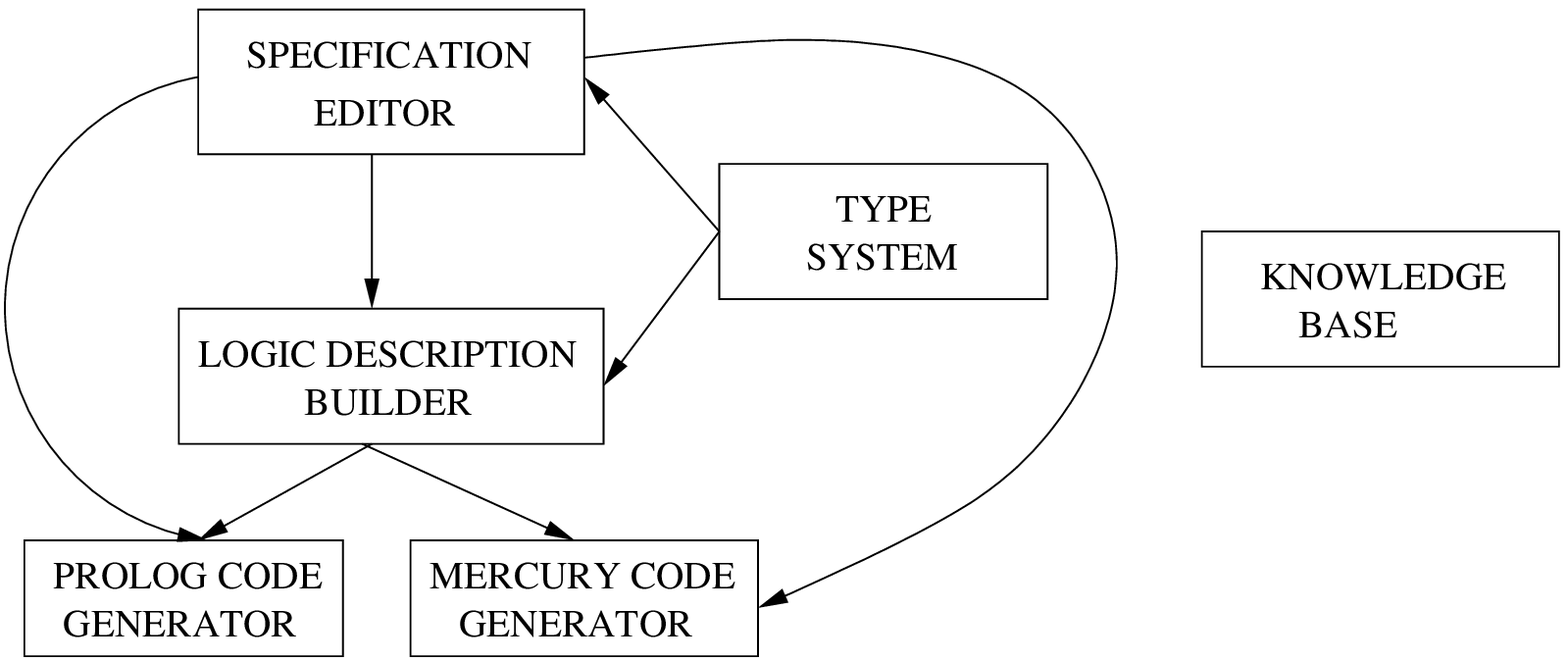}
\caption{Component Architecture of the tool}
\end{center}
\end{figure}

Our tool will be implemented to have two modes of use. An {\em interactive} mode, in which an user interface guides the programmer in all the steps of the development process. The programmer introduces all the needed data through that interface and obtain the documented source code of the built logic procedure. By means of a {\em non interactive} mode, the programmer feeds the tool with heuristic information stored in external files. The tool generates automatically the source code according to the heuristic information given by the user.

Now we will see more details of the tool components.

\subsection{Knowledge Base}
The knowledge base is the main component for the two modes of use of the tool, since all the other components interacts with it.

There exists two kinds of knowledge to be stored in the base. On one hand, a {\em technical knowledge} which includes construction heuristics for typed logic descriptions and static analysis. On the other hand, an {\em object knowledge} which contains all the elements required for the development process: specifications, types and logic programs already built.

When the programmer develops new logic procedures and types, some relevant information about their development are stored in the knowledge base. In the same way, the programmer can adapt the techniques for constructing typed logic descriptions and add new techniques.

\subsection{Specification Editor}
The tool must be able to manipulate the specifications of logic procedures already developed. This is made by the specification editor component.

Our specification format is naturally informal. It reduces the capabilities for automating the program verification, but it is more expressive. An informal specification does not exclude the use of formal notations or methods. This is why we believe that different tools of (semi)formal specifications could interact with this component.

The specification editor will handle all the information about the specifications of logic procedures already developed. The programmer can develop a procedure from the scratch starting with a new specification or he can modify an existing specification to develop a new version of the procedure.

The only automated task is the verification of consistency of the directionalities. If the verification is successful the tool will add the specification to its knowledge base.

\subsection{Type System}
The FOLON environment did not have a type system. It allowed the definition of a type as any arbitrary set of ground terms. One of the novelties of our tool is a separated component for the type system described in a previous section.

The knowledge base will include some built-in types . The programmer will be able to add new types to the knowledge base.

The programmer could manage several implementations of a type, this means, having differents logic descriptions of type-checking for the same data structure. As an example we show two possible implementations of the {\em stack} data type.

\begin{tabular}{l c l}
 & & \\
$stack\_impl1(X)$ & $\Leftrightarrow$ & $list(X)$\\
$stack\_impl2(X)$ & $\Leftrightarrow$ & $X = empty\_stack$\\
 & & $\vee~ X = push(E, S) \wedge stack\_impl2(S)$\\
 & &
\end{tabular}

However, the programmer must be restricted to use only one implementation of a particular type, when he is developing a procedure that uses a type with many implementations.

\subsection{Logic Description Builder}
In the interactive mode, the programmer is guided by means of {\em suggestion mechanisms} on the knowledge base, which represent techniques for constructing typed logic descriptions. The knowledge base could be fed with new techniques.

The techniques that will be available for constructing typed logic descriptions are:

\begin{itemize}
\item Structural Induction
\item Tupling Generalization
\item Generalization by Acummulator Introduction
\end{itemize}

The programmer will also be able to write directly the typed logic description in a file. The tool will parse it and will translate it automatically in a non typed logic description.

\subsection{Code Generators}
One of our goals is to be able to extend the tool with different code generators, one for each possible implementation language. At this moment we are just implementing the code generator for Prolog.

The first subcomponent of a Prolog code generator is a {\em derivator} which receive the non typed logic description and transforms it to a first version of the logic procedure.

This first version is processed by a {\em type-mode analyzer}, which will elminate redundant type-checking literals and will find a permutation of literals in the clauses that satisfies the specified directionalities. The analyzer will use abstract interpretation techniques and will take as starting point the results of the FOLON project about static analysis of types \cite{deboeck} and program verification \cite{rossi}.

If there is no permutation satisfying the directionalities, the tool will suggest two alternatives: either to generate separate versions of the procedure for each directionality, or to change the specification adapting the directionalities.

The programmer will be able to modify manually the order of literals in the clauses or delete type-checking literals. After the analysis, the programmer will receive the source code with some messages about warnings on possible inconsistencies.

The Prolog code generator is enriched with language-dependent optimizers, based on program transformation techniques. Some optimizations like cut introductions can be made according to the specification of directionalities.

For Mercury, the code generator is simpler. The declarations of types and modes are derived from the declaration of types and directionalities in the specification, following a correspondance like the one described in Fig. 2. The generation of code is made directly from the typed logic description, since the Mercury compiler makes itself the analysis of types and modes. Some mechanismes of source-to-source transformation for could be added to get a more optimal code.

\section{Conclusions}

Programming environments have two basic utilities: to facilitate the learning of a programming language and its computation model, and to help programmers to build correct programms in an efficient way.

We have presented a variant of a general methodology for developing logic programs. This variant adds a simple type system and the notion of typed logic description, which give facilities for the development of programs in typed logic languages.

We also present the design of a computational tool that supports our variant of the methodology. The tool is directed to two different publics. In one hand, students and novice programmers who wants to learn the logic programming paradigm with small programs. In the other hand, more experienced programmers who looks for a semiautomatic support on development of middle-scale software systems.

We believe that other techniques of logic program development can be adapted to the methodology. Stepwise enhancement \cite{jain} and synthesis from schemas \cite{flener} could be used at the step of the construction of the typed logic description.

At this moment, we are working on the implementation of a prototype of the tool using the Mercury logic programming language. With the use of Mercury as implementation language we intend to show that logic programming is a good alternative to build CASE tools.

Further work will study the adaptability of different techniques for logic program development in the tool, as well as the development of code generators for Mercury and other declarative languages.

\paragraph{Acknowledgements.}

The first author thanks to Vladimir T\'amara and Luis Quesada, as well as the anonymous reviewers, for their remarks and comments on first drafts of this paper.

\bibliographystyle{plain}

\end{document}